\documentclass[prl,twocolumn,aps]{revtex4-2}
\usepackage{graphicx}
\usepackage{amsfonts}
\usepackage{hyperref}
\begin{document}
\newcommand{\bra}{\langle}
\newcommand{\ket}{\rangle}
\newcommand{\al}{\alpha}
\newcommand{\be}{\beta}
\newcommand{\ga}{\gamma}
\newcommand{\de}{\delta}
\newcommand{\D}{\Delta}
\newcommand{\ep}{\epsilon}
\newcommand{\varep}{\varepsilon}
\newcommand{\e}{\eta}
\renewcommand{\th}{\theta}
\newcommand{\Th}{\Theta}
\newcommand{\la}{\lambda}
\newcommand{\La}{\Lambda}
\newcommand{\Ga}{\Gamma}
\newcommand{\m}{\mu}
\newcommand{\n}{\nu}
\renewcommand{\r}{\rho}
\newcommand{\si}{\sigma}
\newcommand{\Si}{\Sigma}
\newcommand{\ta}{\tau}
\newcommand{\vp}{\varphi}
\newcommand{\p}{\phi}
\renewcommand{\c}{\chi}
\newcommand{\ps}{\psi}
\renewcommand{\o}{\omega}
\renewcommand{\O}{\Omega}
\newcommand{\OO}{{\cal O}}
\newcommand{\C}{{\cal C}}
\newcommand{\pa}{\partial}
\newcommand{\beq}{\begin{equation}}
\newcommand{\eeq}{\end{equation}}
\newcommand{\mc}{\mathcal}
\newcommand{\mb}{\mathbb}
\newcommand{\mycomment}[1]{}
\newcommand{\wh}[1]{\widehat{#1}}
\newcommand{\fk}{f_\textup{bulk}}
\newcommand{\fy}{f_\textup{bdy}}
\newcommand{\yi}{y_i}
\newcommand{\yl}{y_L}
\newcommand{\yr}{y_R}

\newcommand{\nn}{\nonumber}
\newcommand{\Sl}{\sum\limits}
\newcommand{\blue}{\color{blue}}
\newcommand{\red}{\color{red}}
\newcommand{\green}{\color{green}}
\newcommand{\black}{\color{black}}

\def\msout#1{\textrm{\sout{#1}}}
\def\mxout#1{\textrm{\xout{#1}}}

 \def\be{\begin{equation}}
\def\ee{\end{equation}}
\def\bea{\begin{eqnarray}}
\def\eea{\end{eqnarray}}
\def\nn{\nonumber}

\def\tr{{\mbox{tr}}}
\def\Atr{{\mbox{Tr}}}

\def\a{\alpha}
\def\b{\beta}
\def\g{\gamma}
\def\d{\delta}
\def\lam{\lambda}
\def\u{\mu}
\def\v{\nu}
\def\r{\rho}
\def\t{\tau}
\def\z{\zeta}
\def\s{\sigma}
\def\th{\theta}

\def\te{\tilde{e}}
\def\tK{\tilde{K}}
\def\tB{\tilde{B}}
\def\htK{\hat{\tilde{K}}}

\def\Qh{\hat{Q}}
\def\baret{\overline{\eta}}
\def\homega{{\hat{\omega}}}
\def\bpsi{{\overline{\psi}}}
\def\wtau{{\widetilde{\tau}}}
\def\bth{{\overline{\theta}}}
\def\blam{{\overline{\lambda}}}

\def\da{{\dot{\a}}}
\def\db{{\dot{\b}}}
\def\dg{{\dot{\g}}}

\def\bj{{\overline{j}}}
\def\bk{{\overline{k}}}
\def\bz{{\overline{z}}}
\def\sa{{\hat{a}}}
\def\sb{{\hat{b}}}
\def\sc{{\hat{c}}}
\def\wa{{\tilde{a}}}
\def\wb{{\tilde{b}}}
\def\wc{{\tilde{c}}}
\def\oa{{\overline{a}}}
\def\ob{{\overline{b}}}
\def\oc{{\overline{c}}}
\def\od{{\overline{d}}}

 \def\CA{{\cal A}}
 \def\CC{{\cal C}}
 \def\CF{{\cal F}}
 \def\CI{{\cal I}}
 \def\cJ{{\cal J}}
 \def\tJ{{\tilde J}}
 \def\tcJ{{\tilde\cal J}} 
 \def\CO{{\cal O}}
 \def\o{{\rm ord}}
 \def\Ph{{\Phi }}
 \def\L{{\Lambda}}
 \def\CN{{\cal N}}
 \def\p{\partial}
 \def\pslash{\p \llap{/}}
 \def\Dslash{D \llap{/}}
 \def\apm{{\a^{\prime}}}
 \def\r{\rightarrow}
\def\ts{\tilde s}
\def\tu{\tilde u}
 \def\BR{\IR}
 \def\BZ{\IZ}
 \def\BC{\IC}
 \def\BM{\QM}
 \def\BP{\IP}
 \def\BH{\QH}
 \def\BX{\QX}
 \def\sym#1{{{\rm SYM}} _{#1 +1}}
 \def\imp{$\Rightarrow$}
 \def\IZ{\relax\ifmmode\mathchoice
 {\hbox{\cmss Z\kern-.4em Z}}{\hbox{\cmss Z\kern-.4em Z}}
 {\lower.9pt\hbox{\cmsss Z\kern-.4em Z}}
 {\lower1.2pt\hbox{\cmsss Z\kern-.4em Z}}\else{\cmss Z\kern-.4em Z}\fi}
 \def\IB{\relax{\rm I\kern-.18em B}}
 \def\IC{{\relax\hbox{$\inbar\kern-.3em{\rm C}$}}}
 \def\Ic{{\relax\hbox{$\inbar\kern-.22em{\rm c}$}}}
 \def\ID{\relax{\rm I\kern-.18em D}}
 \def\IE{\relax{\rm I\kern-.18em E}}
 \def\IF{\relax{\rm I\kern-.18em F}}
 \def\IG{\relax\hbox{$\inbar\kern-.3em{\rm G}$}}
 \def\IGa{\relax\hbox{${\rm I}\kern-.18em\Gamma$}}
 \def\IH{\relax{\rm I\kern-.18em H}}
 \def\II{\relax{\rm I\kern-.18em I}}
 \def\IK{\relax{\rm I\kern-.18em K}}
 \def\IP{\relax{\rm I\kern-.18em P}}

\def\Tr{{\rm Tr}}
 \font\cmss=cmss10 \font\cmsss=cmss10 at 7pt
 \def\IR{\relax{\rm I\kern-.18em R}}

\def\wdg{{\wedge}}

\newcommand\ev[1]{{\langle {#1}\rangle}}
\newcommand\SUSY[1]{{{\cal N} = {#1}}}
\newcommand\diag[1]{{\mbox{diag}({#1})}}
\newcommand\com[2]{{\left\lbrack {#1}, {#2}\right\rbrack}}

\newcommand\px[1]{{\partial_{#1}}}
\newcommand\qx[1]{{\partial^{#1}}}

\newcommand\rep[1]{{\bf {#1}}}

\def\gam{{\widetilde{\gamma}}} 
\def\sig{{\sigma}} 
\def\hsig{{\hat{\sigma}}} 

\def\eps{{\epsilon}} 

\def\bZ{{\overline{Z}}}
\def\BR{{\mathbb R}}
\def\Lag{{\cal L}}
\def\cO{{\cal O}}
\def\cH{{\cal H}}
\def\wcO{{\widetilde{\cal O}}}
\def\vL{{\vec{L}}}
\def\vx{{\vec{x}}}
\def\vy{{\vec{y}}}

\def\vLf{{\vec{\lambda}}}

\newcommand\cvL[1]{{L^{#1}}} 

\def\npsi{{\psi^{(inv)}}}
\def\bnpsi{{\bpsi^{(inv)}}}
\def\wpsi{{\widetilde{\psi}}}
\def\bwpsi{{\overline{\widetilde{\psi}}}}
\def\hpsi{{\hat{\psi}}}
\def\bhpsi{{\overline{\hat{\psi}}}}

\newcommand\nDF[1]{{{D^F}_{#1}}}
\def\wA{{\widetilde{A}}}
\def\wF{{\widetilde{F}}}
\def\wJ{{\widetilde{J}}}
\def\hA{{\hat{A}}}
\def\hF{{\hat{F}}}
\def\hJ{{\hat{J}}}
\def\MapL{{\Upsilon}}
\def\hR{{\hat{R}}}
\def\hL{{\hat{L}}}

\def\cpl{{\lambda}}  
\def\mcr{{\mathcal{R}}}
\def\xv{{\vec{x}}}
\def\xvt{{\vec{x}^{\top}}}
\def\nv{{\hat{n}}}
\def\nvt{{\hat{n}^{\top}}}
\def\hM{{M}}
\def\hgM{{\widetilde{M}}}
\def\utr{{\mbox{tr}}}

\newcommand{\fig}[1]{fig.\ (\ref{#1})}
\def\dd{\mbox{d}}
\def\ddd{\mbox{\sm d}}
\def\o{\omega}
\def\bra{\langle}
\def\ket{\rangle}
\def\a{\alpha}
\def\b{\beta}
\def\bb{{\bar{\beta}}}
\def\d{\delta}
\def\dd{\partial}
\def\D{\Delta}
\def\LL{\triangle}
\def\g{\gamma}
\def\G{\Gamma}
\def\e{\epsilon}
\def\ve{\varepsilon}
\def\et{\eta}
\def\f{\phi}
\def\F{\Phi}
\def\vf{\varphi}
\def\k{\kappa}
\def\l{\lambda}
\def\L{\Lambda}
\def\m{\mu}
\def\n{\nu}
\def\s{\sigma}
\def\S{\Sigma}
\def\o{\omega}
\def\p{\pi}
\def\r{\rho}
\def\t{\tau}
\def\bt{\bar{\tau}}
\def\th{\theta}
\def\vt{\vartheta}
\def\ra{\rightarrow}
\def\la{\leftarrow}
\def\pa{\partial}
\def\ov{\overline}
\def\Pl{s_{\sm{Pl}}}
\def\tr{\tilde R}
\def\td{\tilde d}
\def\gmn{g_{\mu \nu}}
\def\DO{\D_2}
\def\O{{\cal O}}

\newcommand{\ti}[1]{\tilde{#1}}
\renewcommand{\^}[1]{\hat{#1}}
\newcommand{\sm}[1]{\mbox{\scriptsize #1}}
\newcommand{\tn}[1]{\mbox{\tiny #1}}
\renewcommand{\@}[1]{\sqrt{#1}}
\renewcommand{\le}[1]{\label{#1}\end{eqnarray}}
\newcommand{\eq}[1]{(\ref{#1})}
\def\nn{\nonumber\\}
\def\nm{\nonumber}
\newcommand{\rf}[1]{\cite{ref:#1}}
\newcommand{\rr}[1]{\bibitem{ref:#1}}
\def\qu{\ {\buildrel{\displaystyle ?} \over =}\ }
\def\smqu{\ {\buildrel ?\over =}\ }
\def\ffract#1#2{\raise .35 em\hbox{$\scriptstyle#1$}\kern-.25em/
\kern-.2em\lower .22 em \hbox{$\scriptstyle#2$}}
\def\GN{G_{\mbox{\tn N}}}
\def\lPl{s_{\mbox{\tn{Pl}}}}
\def\fn{f_{(0)}}
\def\fe{f_{(1)}}
\def\ft{f_{(2)}}
\def\fd{f_{(3)}}
\def\Ric{{\mbox{Ric}}}
\def\Rie{{\mbox{Rie}}}
\def\Ein{{\mbox{Ein}}}
\def\nl{\newline}
\def\cl{\textcolor}
\def\clb{\colorbox}
\def\na{\nabla}
\def\half{{1\over2}\,}
\def\nonu{\nonumber \\{}}
\def\da{\dot{a}}
\def\db{\dot{b}}
\def\dc{\dot{c}}
\def\tg{\tilde{\g}}
\def\bF{\bar{F}}
\newcommand{\N}{\mathbb{N}}
\newcommand{\Z}{\mathbb{Z}}


\def\vwsX{{\bf X}} 
\def\dvwsX{{\dot{\bf X}}} 
\newcommand\vxmod[1]{{{\bf X}_{#1}}} 
\newcommand\dvxmod[1]{{\dot{{\bf X}}_{#1}}} 

\def\vwsP{{\bf \Pi}} 

\def\wsX{{X}} 
\def\wsZ{{Z}} 
\def\bwsZ{{\overline{Z}}} 
\def\wsPsi{{\Psi}} 
\def\bwsPsi{{\overline{\Psi}}} 

\def\twsZ{{\widetilde{Z}}} 
\def\twS{{\widetilde{S}}} 
\def\btwsZ{{\overline{\widetilde{Z}}}} 
\def\twsPsi{{\widetilde{\Psi}}} 
\def\btwsPsi{{\overline{\widetilde{\Psi}}}} 
\def\tV{{\widetilde{V}}}
\def\talp{{{\tilde{\a}}'}} 

\def\hLf{{\hat{\lambda}}}
\def\hL{{\hat{L}}}
\def\bh{\bar{h}}
\def\bq{\bar{q}}

\newdimen\tableauside\tableauside=1.0ex
\newdimen\tableaurule\tableaurule=0.4pt
\newdimen\tableaustep
\def\phantomhrule#1{\hbox{\vbox to0pt{\hrule height\tableaurule width#1\vss}}}
\def\phantomvrule#1{\vbox{\hbox to0pt{\vrule width\tableaurule height#1\hss}}}
\def\sqr{\vbox{%
  \phantomhrule\tableaustep
  \hbox{\phantomvrule\tableaustep\kern\tableaustep\phantomvrule\tableaustep}%
  \hbox{\vbox{\phantomhrule\tableauside}\kern-\tableaurule}}}
\def\squares#1{\hbox{\count0=#1\noindent\loop\sqr
  \advance\count0 by-1 \ifnum\count0>0\repeat}}
\def\tableau#1{\vcenter{\offinterlineskip
  \tableaustep=\tableauside\advance\tableaustep by-\tableaurule
  \kern\normallineskip\hbox
    {\kern\normallineskip\vbox
      {\gettableau#1 0 }%
     \kern\normallineskip\kern\tableaurule}%
  \kern\normallineskip\kern\tableaurule}}
\def\gettableau#1 {\ifnum#1=0\let\next=\null\else
  \squares{#1}\let\next=\gettableau\fi\next}

\tableauside=1.0ex
\tableaurule=0.4pt



\newcommand{\eeqn}{\end{eqnarray}}
\newcommand{\ack}[1]{{\bf Pfft! #1}}
\newcommand{\osigma}{\overline{\sigma}}
\newcommand{\orho}{\overline{\rho}}
\newcommand{\myfig}[3]{
	\begin{figure}[ht]
	\centering
	\includegraphics[width=#2cm]{#1}\caption{#3}\label{fig:#1}
	\end{figure}
	}
\newcommand{\littlefig}[2]{
	\includegraphics[width=#2cm]{#1}}

\title{
Modular Bootstrap, Elliptic Points, and Quantum Gravity}
\author{ Ferdinando~Gliozzi}
\affiliation{ Dipartimento di Fisica, Universit\`a di Torino
and Istituto Nazionale di Fisica Nucleare - Sezione di Torino
Via P. Giuria 1 I-10125 Torino, Italy.
}


\date{\today}
\begin{abstract}
  The modular bootstrap program for two-dimensional conformal field theories could be seen as a systematic exploration of the physical consequences
  of consistency conditions at the elliptic points and at the cusp of their
  torus partition function. The study at $\tau=i$, the elliptic point stabilized by the modular inversion $S$, was initiated by Hellerman, who found a general upper bound for the most relevant scaling dimension $\Delta$. Likewise,
  analyticity at $\tau=i\infty$, the cusp stabilized by the modular translation $T$, yields an upper bound on the twist gap. Here we study
  consistency conditions at $\tau=\exp[2i\pi/3]$, the elliptic point stabilized by $S\,T$. We find a much stronger upper bound in the large-$c$ limit, namely
  $\Delta<\frac{c-1}{12}+0.092$, which is very close to
  the minimal mass threshold of  the BTZ black holes in the gravity
  dual of $AdS_3/CFT_2$ correspondence.  
\end{abstract}
\pacs{}

\maketitle

\section{Introduction}
It is still an open question as to whether three-dimensional pure gravity exists as a quantum theory.
In the case of negative cosmological constant, according to holographic
duality\cite{Maldacena:1997re}, solving pure quantum gravity means finding the two-dimensional conformal field theory (CFT) defined on the boundary of the asymptotically anti de Sitter (AdS) spacetime. At the classical level, pure $3d$ gravity is ``trivial'' in the sense that there are no gravitational waves; its degrees of freedom correspond to multitrace composites of the stress-tensor which map to the Virasoro module of the identity of the  CFT.

As a consequence, if the degrees of freedom were only those  of the identity
module, pure gravity would not admit a quantum completion. The reason is very simple: the partition function of the  CFT on the toroidal boundary is modular invariant, while the character of the identity module, as well as any other single Virasoro character, is not. Thus modular invariance of the boundary theory implies additional degrees of freedom in the bulk.

What is the meaning of modular invariance on the gravity side? In a
quantum approach to gravity one expects to sum  over different
topologies of spacetime with fixed asymptotic boundary conditions. It
is widely believed that modular invariance arises from the sum over
saddle points of the gravitational path integral
\cite{Dijkgraaf:2000fq,Manschot:2007ha,Keller:2014xba}. One such
geometry is thermal AdS with periodic Euclidean time. Others
correspond to the Euclidean black holes discovered by Ba\~nados,
Teitelboim and Zanelli (BTZ) in any three-dimensional gravity with
negative cosmological constant \cite{Banados:1992wn}. Thus BTZ black
holes are necessary degrees of freedom for a quantum description of
pure gravity in asymptotic $AdS_3$ spacetimes. Are these enough? BTZ
black holes can exist only above some mass threshold that in the
large-$c$ limit is holographically dual to a CFT primary of scaling
dimension 
$\Delta_{BTZ} =\frac {c-1}{12}$, where $c$ is the central charge.

Primary operators with $\Delta<\Delta_{\rm BTZ}$ can not be interpreted as black holes; they correspond to a new kind of matter  \cite{comments}.  Therefore, proving that a primary with $\Delta<\Delta_{\rm BTZ}$ is necessary for a consistent CFT at the boundary would be sufficient to argue that pure quantum gravity does not exist.

Modular invariance of the partition function constrains the possible spectra of $2d$ CFTs \cite{Cardy:1986ie}. In particular, as first pointed out by Hellerman \cite{Hellerman:2009bu}, for general unitary $2d$ CFTs with $c>1$ it is possible to find an upper bound for the allowed scaling dimension, $\Delta_0$, of the first non-trivial primary. 
Hellerman rigorously established the inequality
$\D_0< \frac c6 +0.4737$. This bound has since been improved numerically as well as analytically in various ways
\cite{Friedan:2013cba,Hartman:2014oaa,Qualls:2014oea,Collier:2016cls,
  Afkhami-Jeddi:2019zci,Hartman:2019pcd,
    Benjamin:2019stq,Maxfield:2019hdt}, in particular using the linear programming method introduced in \cite{Rattazzi:2008pe}.
So far, in the large-$c$ limit the best analytic bound is \cite{Hartman:2019pcd}
$\Delta_0\leq c/8.503$, while the numerical upper bound, obtained by extrapolating large-$c$ data, is \cite{Afkhami-Jeddi:2019zci} $\Delta_0\leq c/9.08$.
Other bounds can be found by assuming complete factorization of the partition function into a holomorphic and an antiholomorphic part \cite{Witten:2007kt}. In this case, the CFT includes an infinite set of conserved higher spin currents, while it is widely believed that pure gravity is dual to a CFT in which the only conserved currents are those generated by the stress tensor.

$PSL(2,\Z)$, the modular group of the torus, is the group of linear fractional transformations acting on the modular parameter $\tau\in H_+$ ($H_+$ is the upper half plane) and generated by the inversion $S$ and the translation $T$, satisfying $S^2=(ST)^3=1$.
One of the reasons why the modular invariance of the partition function $Z$ is so constraining is that $PSL(2,\Z)$ does not act freely on $H_+$: there are points of
$H_+$ which are left invariant under the action of some non-trivial subgroup of
$PSL(2,\Z)$. The partition function is a smooth function of $\t$  only if it fulfills certain consistency conditions at these special points. $PSL(2,\Z)$ admits three points of this kind. So far, only two of them have been extensively explored in modular bootstrap studies. The third one, left invariant by $ST$, has
been considered so far only by  Qualls \cite{Qualls:2014oea}, who wrote a set
of consistency equations, under the assumption that there are partition functions in which only even spin primaries contribute.

In this paper we considerably enlarge the set of equations considered by
Qualls. These give rise to a wealth of information about infinitely many primary
operators for any $2d$ CFT with central charge $c>1$. Equations
(\ref{firsteqs}) show some examples of them, written
as simple sum rules involving both the spins and the scaling dimensions of
primary operators.   

The large-$c$ limit of these sum rules is especially interesting because
unitary CFTs with large $c$  are holographically dual to quantum gravity in
asymptotically AdS spacetimes. In this limit we derive from them a tighter
upper bound on the allowed  scaling dimensions of the first    
non-trivial primary, namely,  
\beq\label{newbound}
\D_0<\frac{c-1}{12}+\frac1{2\sqrt{3}\,\pi}\,,
\eeq
which is  far  stronger than those found to date, and is valid under the same
assumptions of \cite{Hellerman:2009bu}, with the further specification that
the scaling dimension of the first odd-spin primary must lie below $\D_0$.

It is worth stressing that this bound is remarkably close to the threshold
$\D_{BTZ}=\frac{c-1}{12}$
which on the gravitational side constrains the minimal mass of black holes.
In addition, our derivation shows that this upper bound is not an extremum,
suggesting it should be possible to improve it further. Even a modest
improvement  could  push $\D_0$
down the BTZ threshold, implying that 
pure Einstein gravity in $AdS_3$ do not exist as a quantum theory.

  \section{Constraints from modular invariance}
  Modular invariance of the partition function is guaranteed to be a universal property of any physically meaningful theory formulated on a torus, since modular transformations correspond to changes of the  basis on $\Z+\t\Z$,
  its period lattice, while the physics can not depend on the choice of this basis. It reads
  \beq
  Z(\tau,\bar{\tau})=Z\left(\frac{a\tau+b}{c\tau+d},
  \frac{a\bar{\tau}+b}{c\bt+d}\right),\left(\matrix{a &b \cr c &d}\right)\in PSL(2,\Z).
  \label{zinvariance}
  \eeq
 $\t$  and $\bt$ may be considered as two independent complex variables with $\t\in H_+$ and
  $\bt\in H_-$, where $H_+$ and $H_-$ are the upper and the lower half planes.
  $Z(\t,\bt)$ becomes the partition function on a torus of modular parameter $\t$ when $\bt$ is chosen to be the complex conjugate of $\t$, but equation (\ref{zinvariance}) is more general. If the theory on the torus is conformally
  invariant, we can expand $Z$ as a sum over all contributing states
  \beq
  Z(\t,\bt)=\sum_{h,\bh}q^{h-\frac{c}{24}}\bq^{\bh-\frac{c}{24}},\,\,\,q=e^{2i\pi\t},\bq=e^{-2i\pi\bt},
\label{qexpansion}
  \eeq
  where $\D=h+\bh$ are the scaling dimensions of the states and $j=h-\bh$ their spins. For the sake of simplicity we have taken the same central charge $c$ for the left and the right Virasoro algebras. There is a unique vacuum  with $h=\bh=0$. If one further assumes the theory unitary with a discrete spectrum, then equation (\ref{qexpansion}) implies that $Z$ is a holomorphic function on
  $H_+\times H_-$.

  The  partition function of any physical theory formulated on a torus
  is assumed to be a smooth function on the fundamental domain 
   $H_+/PSL(2,\Z)$ (or  $H_-/PSL(2,\Z)$). We can then apply a property
  which is at the core of  modular bootstrap
  \cite{Cardy:1986ie,Hellerman:2009bu}: A smooth function on the fundamental domain  lifts to a smooth function on its covering space $H_+$ (or $H_-$) if and only if it satisfies certain consistency conditions on its derivatives at  the elliptic points and at the cusp, i.e. the special points of the fundamental region which are invariant under the action of some non-trivial subgroup of $PSL(2,\Z)$.

  As already mentioned in the Introduction, the torus modular group
  admits three points of this kind. The cusp at $\t=i\infty$ is stabilized
  by the subgroup generated by $T:\,\t\rightarrow \t+1$. The $\Z_2$ elliptic point
  at $\t=i$ is stabilized  by $S:\,\t\rightarrow -1/\t$. The elliptic point
  at $\t=e^{2i\pi/3}$ is stabilized by $ST:\,\t\rightarrow\,\frac{-1}{\t+1}$,
  the generator of a $\Z_3$ subgroup of $PSL(2,\Z)$.

  Invariance of the partition function under $T$ implies integer spins, i.e.  $j=h-\bh\in\Z$, and  analyticity at the cusp $\t=i\infty$ yields, for $c>1$,
  the upper bound $\D-\vert j\vert\leq\frac{c-1}{12}$ on the twist gap \cite{Collier:2016cls}. Consistency conditions at $\t=i$ demand \cite{Hellerman:2009bu}
  \beq
  \left(\t\partial_{\t}\right)^m \left(\bt\partial_{\bt}\right)^n Z(\t,\bt)
  \vert_{\t=-\bt=i}=0\,\,{\rm for}\,m+n\,{\rm odd}\,.
  \label{hellerman}
  \eeq
  This infinite set of homogeneous linear equations, dubbed modular bootstrap,
  yields the Hellerman upper bound on the scaling dimensions of the lightest primary  operator and its refinements described in the Introduction.

  What are the further constraints dictated by the  consistency conditions 
  at the  point stabilized by $ST$ ? It suffices to take arbitrary derivatives of the identity
  \beq
  Z(\t,\bt)=Z\left(\frac{-1}{\t+1},\frac{-1}{\bt+1}\right)\,,
  \label{identity}
  \eeq
  and evaluate them at  $\t=e^{\frac{2i\pi}3}\equiv\rho$.  We obtain
  \beq
  \pa_\t^n Z\vert_{\t=\rho}-\sum_{m=1}^n\rho^ {m+n} \frac{n!}{m!}\left(\matrix{n-1
    \cr m-1}\right)\pa_\t^m Z\vert_{\t=\rho}=0\,,
  \label{newbootstrap}
  \eeq
  and identical equations for $\bt$. Since $\rho^{3\,k}=1$, iterating the above equations shows that a derivative of arbitrary order in $\t$ (or in $\bt$) at $\t=\rho$ (or $\bt=\bar{\rho}$) can be expressed as a linear combination of derivatives of lower order in multiples of 3. These are the equations to be added to (\ref{hellerman}) to complete the modular bootstrap program.
  
  In order to
  keep $Z$  real it is convenient to parametrize $\t$ and $\bt$ as
  \beq
  \t=-\frac12+i\,\frac{\b}{2\pi}\,\,,\, \bt=-\frac12-i\,\frac{\bb}{2\pi}\,,
  \eeq
  where $\b$ and $\bb$ are two independent real variables, so we
  can set $q=-e^{-\b}$ and $\bq=-e^{-\bb}$ in (\ref{qexpansion}), thus obtaining a real
  partition function. Terms with even  $j=h-\bh$ are positive while
  those with odd $j$  are negative. The $\Z_3$ elliptic point corresponds
  to $\b=\bb=\sqrt{3}\,\pi\equiv\b_c$.

  The first few equations  are, more explicitly,
  \bea\label{equations}
  \pa_\b Z\vert_{\b=\bb=\b_c}=0\,\,,\,\,\,\,\,
  \left( \pa_\b^4+\frac{2\sqrt{3}}\pi \pa_\b^3\right) Z\vert_{\b=\bb=\b_c}&=&0\,,\cr
 \pa_\b^2 Z\vert_{\b=\bb=\b_c}=0\,,\,\,\,\,\,\,\,\,\,\,
 \left( \pa_\b^5-\frac{10}{\pi^2} \pa_\b^3\right) Z\vert_{\b=\bb=\b_c}&=&0\,,
 \,\,\,\,\cr
 \left( \pa_\b^7+\frac{525}{\pi^4}
   \pa_\b^3+\frac{7\sqrt{3}}{\pi}\pa_\b^6\right)
 Z\vert_{\b=\bb=\b_c}&=&0\,,\cr 
\left(\pa_\b^2\pa_\bb+\frac1{\pi\sqrt{3}}\pa_\b\pa_\bb\right)
Z\vert_{\b=\bb=\b_c}&=&0\,,\cr
\left(\pa_\b^3\pa_\bb-\frac1{2\pi^2}\pa_\b\pa_\bb\right)
Z\vert_{\b=\bb=\b_c}&=&0\,,
 \eea
 and identical equations for $\bb$. One can check these identities by applying them for instance to the modular invariant
 \beq
 \sqrt{\t-\bt}\,\eta(\t)\,\eta(-\bt)=Z_b^{-1}\,,
 \eeq
 where $Z_b$ is the partition function of a free boson and $\eta$ is the Dedekind eta function.

 In order to obtain useful information on the Virasoro primary spectrum,
 we have to separate primaries from descendants in the sum of states
 (\ref{qexpansion}). $Z(\t,\bt)$ can be expanded in Virasoro characters.
 If $c>1$ and the theory is unitary, the modules of the Virasoro algebra are the identity degenerate module $\chi_0(q)$ and a continuous family of non-degenerate modules $\chi_A(q)$ labeled by a positive conformal weight $h_A$
 \beq
 \chi_0(q)=\frac{q^{-\frac{c-1}{24}}}{\eta(\t)}(1-q)\,,\,\,\,\,\, \chi_A(q)=
   \frac{q^{h_A-\frac{c-1}{24}}}{\eta(\t)}\,.
   \eeq
   Assuming discreteness of the spectrum and no conserved currents
   beyond those of the Virasoro algebra  yields the expansion
   $Z=[0_0]+\sum_AN_A[\D_{j_A}]$, or more explicitly
   \beq
   Z(\t,\bt)=\chi_0(q)\chi_0(\bq)+\sum_AN_A\,\chi_A(q)\,\chi_A(\bq)\,,
   \label{chexpansion}
   \eeq
   where the multiplicity $N_A$ is a non-negative integer.

   In order to obtain the promised upper bound it suffices to apply to
   such an expansion the first few identities (\ref{equations}). We get rid of the $\eta$ function and its derivatives by considering the modular-invariant combination
   $Z(\t,\bt)/Z_b=Z_{vac}+\sum_A N_A\,Z_A$, with
   \bea
   Z_{vac}&=&\sqrt{\b+\bb}\,\, e^{\b\frac{c-1}{24}} e^{\bb\frac{c-1}{24}}\,
   (1+e^{-\b})(1+e^{-\bb})\,,\cr
Z_A&=&\sqrt{\b+\bb}\,\, e^{\b\frac{c-1}{24}} e^{\bb\frac{c-1}{24}}\,e^{-\b h_A}e^{-\bb \bh_A}\,.
\eea
When applying (\ref{equations}) it turns out that the scaling dimensions $\D_A=h_A+\bh_A$ always appear in the combination $\D_A-\D_+$ with
\beq
\D_+=\frac{c-1}{12}+\frac1{2\sqrt{3}\,\pi}\,.
\label{deltaplus}
\eeq

It is convenient to organize  equations  (\ref{equations}) in terms
of polynomials of the two differential operators $\pa_\b+\pa_\bb $ and
$\pa_\b -\pa_\bb $. Odd powers of
the latter give trivial identities as a consequence of the symmetry of the
partition function under the transformation  $j_A \rightarrow -j_A$.
The other identities can be written as surprisingly simple sum rules;
the 
first two are \cite{comments2} 
\bea\label{firsteqs}
\sum_A(-1)^{j_A}N_A\, e^{-\b_c\D_A}&&(\D_A-\D_+)=v^2\D_+-uv\,,\cr
\sum_A(-1)^{j_A}N_A\, e^{-\b_c\D_A}&&\left((\D_A-\D_+)^2-
\frac1{6\pi^2}+j^2_A\right)=\cr
-v^2\D_+^2
&&+2uv(\D_+-1)+\frac{v^2}{6\pi^2}\,,
\eea
with $u=2\,e^{-\b_c},\,v=1+e^{-\b_c}$.
The LHS of these sum rules measures the difference in contributions 
of even and odd spins. 
We expect that at large $j$ the two kinds of contributions
cancel, as the density of the states is the same in this limit
\cite{Kusuki:2018wpa,Collier:2018exn,Kusuki:2019gjs}, however we do not use such a cancellation in deriving our upper bound.

We can check the above equations  in some specific model. An instructive example is the partition function $Z_f$ of  8 free fermions with diagonal GSO projection, which saturates the unitarity bound at $c=4$ \cite{Collier:2016cls,Hartman:2019pcd}. The full partition function reads
\beq
Z_f(\t)=\sum_{i=2}^4\frac{\theta_i(\t)^4\,\theta_i(-\bt)^4}{2\,\eta(\t)^4\,\eta(-\bt)^4}\,,
\eeq
where $\theta_i$ denote the Jacobi theta functions. $Z_f$ vanishes for $\t\rightarrow e^{2i\pi/3}$, as it becomes proportional to the Eisenstein series $E_4(\t)$, which is known to have a simple zero at this point, so in this case the
cancellation is complete. In our notations the vanishing of $Z_f$
reads 
\cite{comments1}
\beq
\sum_A(-1)^{j_A}N_A\, e^{-\b_c\D_A}=-v^2\,.
\eeq

We can easily expand $Z_f$ in Virasoro characters. The first few terms
are
\beq\label{decomposition}
Z_f=[0_0]+28 [1_1]+192[1_0]+105[2_2]+1344[2_1]+784[2_0]+\dots
\eeq
This expansion includes conserved currents, i.e. primaries of the form
$[j_j]$ \
that we had excluded in deriving (\ref{firsteqs}). The contribution of
a conserved current to the partition function is
$\chi_j(q)\chi_0(\bq)+\chi_0(q)\chi_j(\bq)$. One can repeat the
calculation that led to (\ref{firsteqs}) and check that the consequent modification
is numerically neglegible. As a matter of fact, the two-level decomposition
(\ref{decomposition}) is enough to give a good
numerical check of the first  equation in (\ref{firsteqs}),
 which involves only first derivatives, while the others require much more
 terms.

Equations (\ref{firsteqs}) and their analogues with higher derivatives become particularly interesting in the large-$c$ limit, where they simply read, for integer $n$,
\beq\label{main}
\sum_A(-1)^{j_A}N_A\, e^{-\b_c\D_A}\left(\frac{\D_A-\D_+}{\D_+}\right)^n=v^2(-1)^{n+1}+O(\frac1c)\,.
\eeq
We arrange the terms of these series following the increasing order of the scaling dimensions of the primaries, i.e. $0<\D_1 \leq\D_2\leq\dots$ and denote by
$S^{(k)}_n$ the partial sum of the first $k$ terms of the $n^{th}$ series.
We assume the convergence of these series. This means that for every real number
$\epsilon>0$ there exist an integer $m_n$ such that for all $k\ge m_n$ we have
$\vert S^{(k)}_n-S_n\vert<\epsilon$, where $S_n\equiv v^2(-1)^{n+1}$ is the sum
of the series. Choosing $\epsilon$  small enough implies that
all the finite sums $S^{(k)}_n$ with $k\ge m_n$ have the same sign of $S_n$.
Thus from the convergence of the first $n_0$ series we obtain the following \underline{\sl exact} $n_0$
inequalities $S^{(m)}_{n=2\ell-1}>0,\,\,S^{(m)}_{n=2\ell}<0$, with
$m={\rm max}(m_n),\,\,(1\leq n\leq n_0)$.
We recast the above inequalities in the form
\be
A_{n}\,\,\bigg\{\matrix{ >\,B_n\,\,\,{\rm if}\,n\,{\rm  odd}\,\cr
  <\,B_n\,\,{\rm if}\, n\,{\rm  even}} \, \,,
  \label{theorem}
\ee
where to simplify the notation we defined $A_{n} =\sum_{i=1}^p w_i\, a_i^{n}$ 
and $B_n=\sum_{j=1}^q z_j\, b_j^{n}$,
with $p+q=m$; the variables $a_1<a_2<\dots<a_p$ and  $b_1<b_2<\dots<b_q$
denote  the even-spin and the odd-spin terms, respectively, and $w_i>0$, $z_j>0$
their multiplicity.

So far we only assumed the convergence of the series.
If we now assume a stronger
hypothesis, namely that the first $n_0$ series, for a suitable $n_0$, fulfill
the condition
\be
m\equiv p+q<n_0\,,
\label{hypothesis}
\ee
we can easily demonstrate that the scaling dimension
of the lowest odd-spin primary lies below $\D_+$, i.e. $b_1<0$.   

     This theorem can be proved by {\sl reductio ad absurdum}, i.e. assuming
     $b_1>0$ we  show that the inequalities (\ref{theorem}) have no solution.

     First, note that some of the $a_i$'s could be negative, so
     their exclusion reinforces the inequalities, and thus we also
     assume  $a_i>0$. 

     It is easy to see that the $A$'s (and the $B$'s) fulfill a linear
     relation of the form $\sum_{j=0}^p\lambda_j\,A_{p+k-j}=0\,,(k=1,2,\dots)$
     which can be rewritten as
     \be
     \sum_{i=1}^p w_i\left(\sum_{j=0}^p\lambda_j\, a_i^{p+k-j}\right)=0\,.
     \label{Aidentity}
     \ee
     To prove this identity, it suffices to point out that the $\lambda_j$'s
     are  the coefficients of the polynomial $ x^k\prod_{j=1}^p(x-a_j)\,
     \equiv\, x^k(\sum_{j=0}^p\lambda_j \,x^{p-j})\,,$ with $\,\lambda_0=1$\,.

     Since all the $a_i$'s are positive, the $\lambda_j$ have
     alternating  signs, i.e. $\lambda_1<0,\lambda_2>0$, and so on, therefore
     applying the inequalities (\ref{theorem}) to the identity
     (\ref{Aidentity})  gives
     \be
     B_{p+k}+\lambda_1\, B_{p+k-1}+\dots \lambda_p \,B_k\,\bigg\{
     \matrix{ <0\,\,{\rm if}\,p+k\,{\rm odd}\cr\,\,\, >0\,\,{\rm if}\,p+k\,
       {\rm even}\,.}
     \label{Bn}
     \ee
     In accordance with the assumption (\ref{hypothesis}) $k$ can
     take the values $k=1,2\dots, q+1$. 

     We  rewrite the LHS of these $q+1$
     inequalities  as $\sum_{j=1}^qy_j\,b_j^k$, with
     $y_j=z_j\prod_{i=1}^p(b_j-a_i)$. Clearly the alternating  signs of
     (\ref{Bn}) are possible only if there are both positive and negative
     $y_j$'s. Following the signs of $y_j$'s  we  split the set of
     the $q$ variables $b_j$'s into two subsets of $p'$ and $q'$ elements
     with $q=p'+q'$. In this way, relabelling the indices, 
     we can recast the $q+1$ inequalities (\ref{Bn}) into the same form as
     (\ref{theorem}), but with a reduced number of variables.

     The iteration of this process terminates
     when we are left with a single variable of type $b$ and  $q'+1=2$
     contradictory inequalities, namely  $b>0$ and $b^2<0$, showing that the set
     (\ref{theorem}) combined with (\ref{hypothesis}) 
     is incompatible with the assumption $b_1>0$, QED.

     On the contrary, allowing $b_1$ to be negative it is easy to find numerical solutions of (\ref{theorem}). Some of them also have negative even-spin terms, suggesting a
  possible improvement of the upper bound.

\section{Discussion and Outlook}

In this paper we pointed out that the partition function of a general CFT on a torus should obey a larger class of equations than those explored so far. The new equations allow us to derive a much stronger  upper bound  for the maximal gap of the first non-trivial primary in the large-$c$ limit. The latter is very close to the mass threshold  of the corresponding BTZ black holes in $AdS_3/CFT_2$ correspondence.

For the sake of completeness, let us recall  how such a threshold emerges in the holographic approach. A BTZ black hole of mass $M$ and spin $j$ in the bulk corresponds to a $(h,\bh)$ primary on the boundary with
\beq\label{correspondence}
h-\frac c{24}=\frac12\left( M\ell+j\right)\,,\,\,\bh-\frac c{24}=
\frac12\left( M\ell-j\right)\,,
\eeq
where $\ell$ is the radius of $AdS_3$, related to the central charge by \cite{Brown:1986nw} $c=\frac{3\ell}{2G}$, with $G$  the Newton constant. Black holes have smooth horizons  only if they fulfill the cosmic censorship condition
$M\ell\ge \vert j\vert$, prohibiting naked singularities in spacetime. Therefore, according to (\ref{correspondence}), a CFT primary corresponds to a black hole in the bulk only if $h,\bh\ge \frac{c}{24}$. One-loop corrections replace $c$ with $c-1$ \cite{Maxfield:2019hdt}. Primaries with $\D<\frac {c-1}{12}$ correspond to objects that do not have a smooth event horizon, the defining feature of  black
holes, thus  should correspond to a new kind of matter in the bulk.

The threshold at $\D=\frac{c-1}{12}$ has  a special meaning even on the CFT side \cite{Friedan:2012jk,Maxfield:2019hdt}.
In particular, in a unitary CFT with boundary, a lower bound is derived for the boundary entropy when {\sl assuming}  $\D_0>\frac{c-1}{12}$ \cite{Friedan:2012jk},
however lower values are possible. Actually some indication that a lower upper bound is necessary for saving unitarity has been recently found \cite{Benjamin:2019stq}, since in the double limit $j\rightarrow\infty,\,\,\bh-\frac{c-1}{24}\rightarrow0$ the density of states of the  vacuum character contribution $\chi_0(q)\chi_0(\bq)$ becomes negative. A possibility to avoid this violation of unitarity is assuming a twist gap no larger than $\frac{c-1}{16}$.

From this perspective, it would be interesting to try to improve our upper
bound (\ref{newbound}). Note that to derive it only some simple properties of the new set of equations (\ref{newbootstrap}) were used. A further study of them
could generate important information on the spectrum of primary states in a general CFT. 

\section{}
The author acknowledges a fruitful correspondence with Slava Rychkov.


\begin{thebibliography}{99}
\bibitem{comments}  This point will be discussed in more detail later.
\bibitem{comments2}{I would like to thank Brian McPeak for pointing out
    to me a mistake in a praevious version of this paper.}
\end{thebibliography}
\end{document}